\documentclass[aps,prb,twocolumn,amsmath,amssymb,superscriptaddress,floatfix]{revtex4-2}
\usepackage{graphicx} 
\usepackage{bm}
\usepackage[usenames]{color}
\bibstyle{apsrev.bib}
\usepackage{amsmath}

\newcommand{\be}{\begin{equation}}
\newcommand{\ee}{\end{equation}}
\newcommand{\beqn}{\begin{eqnarray}}
\newcommand{\eeqn}{\end{eqnarray}}

\begin{document}

\title{Mechanism of defect formation in the quantum annealing of the random transverse-field Ising chain}

\author{R\'obert Juh\'asz}
\email{juhasz.robert@wigner.hun-ren.hu}
\affiliation{HUN-REN Wigner Research Centre for Physics, H-1525 Budapest, P.O.Box 49, Hungary}

\date{\today}

\begin{abstract}
  Based on the strong-disorder renormalization group method, a microscopic mechanism of defect formation in the quantum annealing of the random transverse-field Ising chain is proposed, which represents the annealing process as a gradual aggregation of strongly coupled spin clusters. The ferromagnetic ground state of clusters is either preserved or get excited in pairwise fusions of clusters, depending on the effective annealing rate of the fusion, the latter events being responsible for the appearance of defects in the final state. A consequence of the theory is that, although the Griffiths-McCoy phases surrounding the critical point are gapless, they are still effectively gapped from the point of view of quantum annealing. Thereby we provide an explanation of the finiteness of gap outside of the critical point, which was implicit in an early approach to the problem by Kibble-Zurek scaling [Dziarmaga, Phys. Rev. B {\bf 74}, 064416 (2006)]. Furthermore, by identifying the accessible excitations, we refine the functional form of the vanishing of the gap at the critical point. The defect density in the final state is found to decrease with the annealing time $\tau$, as $n(\tau)\sim \ln^{-2}\left(\frac{\tau}{\ln^2\tau}\right)$ for large $\tau$. In addition to this, our approach gives access also to the density of defects at intermediate times of the annealing process. 
\end{abstract}

\maketitle

\section{Introduction}

The theoretical study of nonequilibrium dynamics of closed quantum systems is a challenging topic, whether it is induced by a sudden quench or by a slow change of the underlying Hamiltonian. The latter situation is motivated by adiabatic quantum computing \cite{albash,hauke,das}, a tool for solving discrete optimization problems \cite{lucas,kadowaki}. Here, a driving field is applied, which is initially dominant, preparing the system in a trivial (typically product) state. Then the strength of the field is slowly turned down to zero, arriving at a classical system, the ground state of which encodes the solution of the optimization problem. The adiabatic theorem ensures that, for infinitely slow driving, the system will remain in its instantaneous ground state throughout the process \cite{amin,albash}.  
A prototypical model of quantum annealing is the ferromagnetic transverse-field Ising chain, which, from the point of view of optimization, is trivial, having a ferromagnetic ground state at zero field, but due to its solvability by mapping to free fermions it is a useful testing ground for concepts in quantum annealing \cite{dziarmaga_prl}. 
As the annealing starts in the paramagnetic phase and ends in the ferromagnetic phase, it must inevitably cross the quantum critical point (QCP) at which the energy gap is zero (in an infinite system). As a consequence, the end state will differ from the ground state i.e. some amount of defect forms, no matter how slow the driving was \cite{kibble,zurek,cz,lukin,king,king2023}.
A general heuristic theory by Kibble and Zurek \cite{kibble,zurek,cz} predicts that, for a linear variation of the control parameter across the transition point, $\Delta(t)\simeq t/\tau$, the error density produced in the annealing process scales with the annealing rate $\tau^{-1}$ as
\be
n(\tau)\sim \tau^{-d\nu/(z\nu+1)},
\ee
where $\nu$ and $z$ are the correlation-length and the dynamical critical exponents of the QCP, respectively, and $d$ is the dimension of the system.   
This conclusion relies on the assumption that the state ceases to adapt to the driving Hamiltonian beyond a point, $\Delta^*$, at which the reciprocal energy gap $1/\epsilon\sim (\Delta^*)^{-z\nu}$ becomes comparable with the time remained to reach the critical point \cite{qKZ}. The correlation length at this ``freezing point'', $\xi_*\sim \Delta_*^{-\nu}$ will be imprinted in the final state, resulting in a density of defects $n\sim \xi_*^{-d}$. This general result was confirmed by exact calculations in the homogeneous transverse-field Ising chain \cite{dziarmaga_prl,grabarits}, where, due to translational invariance, the problem reduces in momentum space to a set of two-level problems treated by Landau and Zener \cite{lz}. 

As opposed to the above ideal case, real optimization problems are inhomogeneous i.e. contain typically random couplings. Owing to the loss of translational invariance, a direct decomposition to Landau-Zener transitions is no longer possible, hence much less is known in this case \cite{dziarmaga_random,caneva,zanca,rmc,mohseni,sadhukhan}. In an early work on this problem, Dziarmaga applied Kibble-Zurek mechanism to the random transverse-field Ising chain (RTIC) \cite{dziarmaga_random}. This approach relies on a result of the strong-disorder renormalization group (SDRG) method by Fisher \cite{fisher}, stating that the dynamical critical exponent at the QCP diverges as $z=\frac{1}{2|\Delta|}$. Inserting this into Kibble-Zurek theory, together with $\nu=2$ \cite{fisher}, leads to that the energy gap of the RTIC would vanish as 
\be
\epsilon \sim |\Delta|^{1/|\Delta|}
\label{epsilon_D}
\ee
when $|\Delta|\to 0$, and to the scaling of error density in the final state of the form
\be
n(\tau)\sim \frac{\ln^2[\ln(\tau/\tau_0)]}{\ln^2(\tau/\tau_0)},
\label{n_D}
\ee
where $\tau_0$ is a constant. 
Numerically exact calculations using free-fermion mapping and time-dependent Bogoliubov theory were to found to be in agreement with the logarithmic scaling of Eq. (\ref{n_D}) \cite{caneva,zanca}.  

Nonetheless, Eq. (\ref{epsilon_D}) is perplexing since, according to this, the energy gap is non-zero outside of the critical point. The QCP is, however, well-known to be surrounded by extended gapless phases, the disordered and the ordered Griffiths-McCoy (GM) phase \cite{GM}. In the former, ferromagnetic domains of arbitrary size can occur, the excitation energy of which has a power-law distribution $P_<(\epsilon)\sim \epsilon^{1/z}$ for $\epsilon\to 0$. The ordered GM phase is related to the disordered one by the duality property of the model, and has the same power law tail of excitation energies.  

In this paper, we go beyond the Kibble-Zurek description and develop a theory of defect production during quantum annealing in the RTIC. Our approach is based on the SDRG treatment of the model and decomposes the annealing process into a series of pairwise fusions of strongly coupled spin clusters. These independent fusion events are themselves Landau-Zener transitions which, depending on the minimal energy gap of fusion, are essentially either adiabatic processes or fast quenches, the latter being responsible for the defects appearing in the final state. This picture also resolves the paradoxical situation concerning the energy gap: although the spectrum is gapless in the GM phases, yet, due to the locality of excitations and the $Z_2$ symmetry of the model, the accessible excitations are those associated with fusion events. This leads to that the GM phases become effectively gapped for the quantum annealing procedure. This effective gap closes at the QCP as $\epsilon_{\Delta}\sim e^{-{\rm const}/|\Delta|}$, which is different from Eq. (\ref{epsilon_D}).
Accordingly, our theory results in a somewhat different form of the scaling of defect density, $n(\tau)\sim \ln^{-2}\left(\frac{\tau/\tau_0}{\ln^2(\tau/\tau_0)}\right)$, which has an additive correction to the leading term $\ln^{-2}(\tau/\tau_0)$ for large $\tau$, rather than a multiplicative one appearing in Eq. (\ref{n_D}). 

The paper is organized as follows. The basic concepts of the microscopic mechanism of the annealing process within the SDRG approach in the case of adiabaticity are presented in Sec. \ref{sdrg}. In Sec. \ref{fusion}, the focus is on fusions of clusters which are potentially the elementary events of defect formation.
In Sec. \ref{nonadiabatic}, the description of the annealing process is generalized to the case of non-adiabaticity. Using the theory developed in previous sections, the effective energy gap in the GM phases is determined in Sec. \ref{sec:gap}, while the scaling of defect density is derived in Sec. \ref{defects}. Finally, results are discussed in Sec. \ref{discussion}. 

\section{SDRG picture of adiabatic annealing}
\label{sdrg}

We consider quantum annealing in the ferromagnetic random transverse-field Ising chain having the time-dependent Hamiltonian on an infinite lattice:
\be
{\cal H}(t) =
-\sum_{i}J_{i}(t)\sigma_i^x \sigma_{i+1}^x-\sum_{i}h_i(t)\sigma_i^z.
\label{H}
\ee
Here $\sigma_i^{x,z}$ are Pauli operators on site $i$ and the positive bonds $J_i(t)$ and fields $h_i(t)$ are changed in time from $t=-\tau$ up to $t=\tau$.
Usually, the couplings are varied linearly as
$J_{i}(t)=\frac{1}{2}\left(1+\frac{t}{\tau}\right)J_{i}$ and  $h_{i}(t)=\frac{1}{2}\left(1-\frac{t}{\tau}\right)h_{i}$,
where the bonds $J_i$ are quenched, independent, identically distributed (i.i.d.) random variables, while the transverse fields $h_i$ are either uniform or quenched i.i.d. variables.  
In the quantitative treatment of the problem, we will use a special protocol in which the bonds are time-independent and are drawn from a power-law distribution in the domain $(0,1)$:
\be
\rho(J)=J^{-1+\theta_0},
\label{Jdist}
\ee
where $\theta_0>0$ controlling the strength of disorder is a free parameter.
The transverse fields are time-dependent,
\be
h_i(t)=h_i^{p(t)},
\label{hpl}
\ee
where $h_i$ are quenched i.i.d. random variables having the same distribution as the bonds, and
\be 
p(t)\equiv\frac{\tau+t}{\tau-t}.
\label{pt}
\ee
This means that the fields are initially uniform, $h_i(-\tau)=1$, thus the process starts at the edge of the disordered Griffiths-McCoy phase (since $J_i<1$), traverses the self-dual critical point at $t=0$, and all fields vanish at the end, $\lim_{t\to\tau}h_i(t)=0$. Note that, at any instant, the distribution of transverse fields is algebraic, which is necessary for our later calculations since in this case, the evolution under SDRG transformation reduces to a one-parameter flow through the exponent $p$.  

\begin{figure}[ht]
\includegraphics[width=8cm]{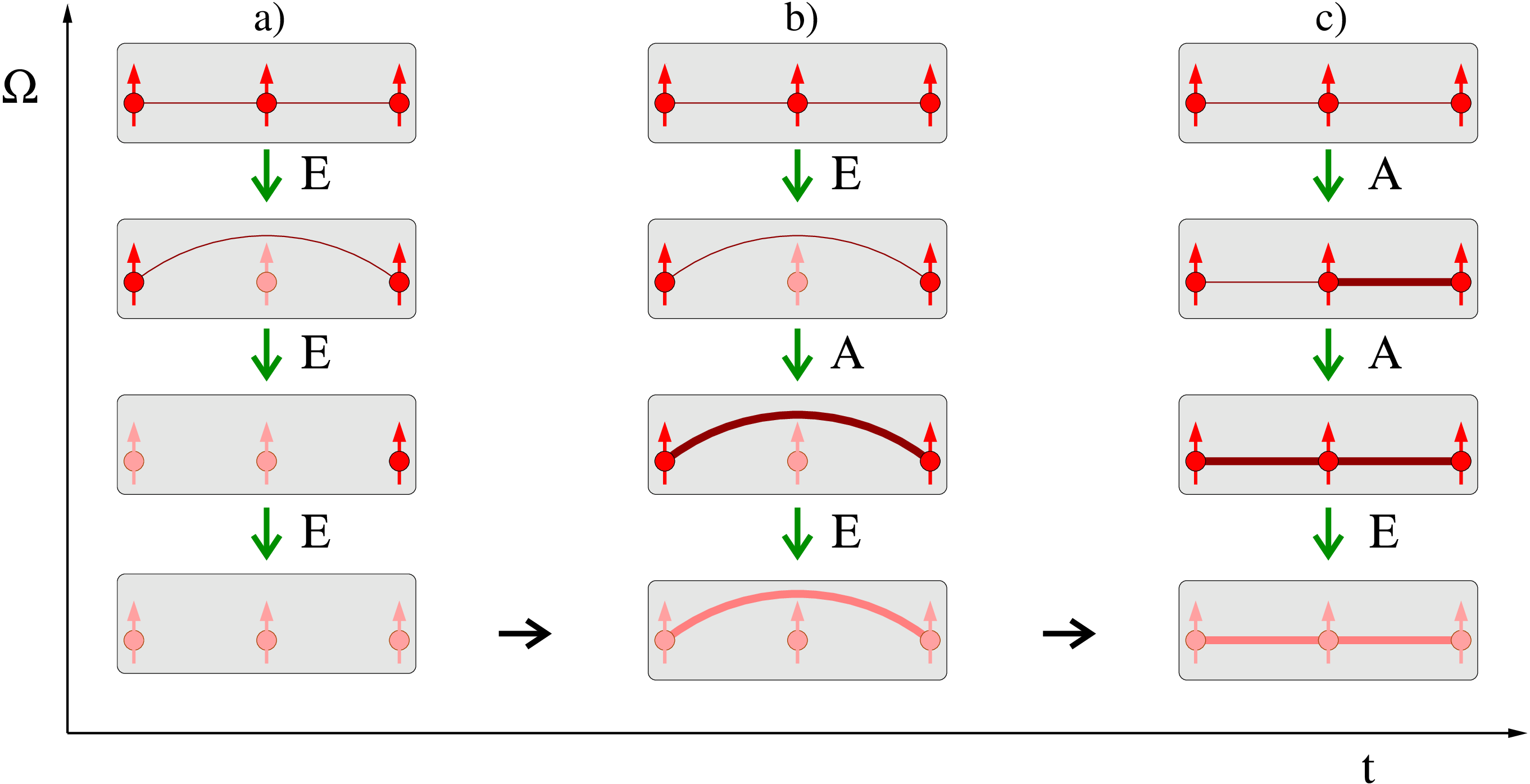}
\caption{\label{scheme}
Renormalization histories of a system of three spins at three different values of $t$. Red (pink) arrows indicate active (decimated) spins. Interactions between spins are symbolized by thin lines, whereas thick lines connect spins aggregated to a common cluster. Down arrows indicate elementary steps of SDRG, which are either aggregations (A) or eliminations (E). The bottom row shows the variation of the cluster set with increasing $t$, which is composed of, in the three depicted cases, three (a), two (b), and one single (c) cluster.
}
\end{figure}

The approximate ground state of the model in Eq. (\ref{H}) at a fixed $t$ can be constructed by the SDRG method \cite{mdh,im} developed for this model by Fisher \cite{fisher}.
In this procedure, blocks of spins containing the largest coupling present in the system are picked, the local Hilbert space is restricted to the low-energy branch of states and effective spin variables are associated with the truncated space by second-order perturbation theory. Applying the block reductions iteratively, the energy scale $\Omega$ and the number of spins in the chain is gradually reduced. The block reductions are of two kinds. If the largest coupling is a transverse field, $\Omega=h_2$, and the neighboring bonds, $J_1$ and $J_2$ are much weaker, spin $2$ freezes to the ground state of the term $-h_2\sigma_2^z$, $|\uparrow_2\rangle$. This spin is then decimated out, and a new interaction term $-\frac{J_{1}J_2}{h_2}\sigma_{1}^x\sigma_{3}^x$ arises between spins neighboring to the decimated one.
If the largest coupling is a bond, $\Omega=J_2$, and the neighboring fields $h_2$ and $h_3$ are much weaker, the local Hilbert space of spin $2$ and $3$ is restricted to the subspace of the low-lying states $|\uparrow_{23}\rangle\equiv\frac{1}{\sqrt{2}}(|\rightarrow_2\rightarrow_{3}\rangle+|\leftarrow_2\leftarrow_{3}\rangle)$ and $|\downarrow_{23}\rangle\equiv\frac{1}{\sqrt{2}}(|\rightarrow_2\rightarrow_3\rangle-|\leftarrow_2\leftarrow_3\rangle)$. The part of the Hamiltonian involved in this truncation is replaced by $\mathcal{H}_{\rm eff}=-J_{1}\sigma_{1}^x\tilde\sigma^x-J_{3}\tilde\sigma^x\sigma_{4}^x-\frac{h_2h_{3}}{J_2}\tilde\sigma^z$, which contains a new spin $\tilde\sigma$ operating on the space spanned by $|\uparrow_{23}\rangle$ and $|\downarrow_{23}\rangle$, which are the ground and excited state of $-\frac{h_2h_{3}}{J_2}\tilde\sigma^z$, respectively.
Applying these steps iteratively results in an aggregation-elimination process of spins to ferromagnetic clusters, in which pairs of clusters are unified to larger clusters by bond decimations ($\Omega=J_i$) and clusters are eliminated (their further growth is terminated) by field decimations ($\Omega=h_i$). For an illustration of this procedure, see Fig \ref{scheme}.
At the fixed point of the SDRG transformation, at which $\Omega=0$, the original spins of the chain are grouped to ferromagnetic  clusters, $C_1,C_2,\dots$ which interact weakly with each other.
The approximate ground state is thus a product $|GS\rangle_{\rm SDRG}=\otimes_n|\uparrow_{C_n}\rangle$ of cluster states, which are themselves maximally entangled Greenberger-Horne-Zeilinger states of the spins belonging to them:
$|\uparrow_{C_n}\rangle=\frac{1}{\sqrt{2}}\left(\otimes_{i\in C_n}|\rightarrow_i\rangle+\otimes_{i\in C_n}|\leftarrow_i\rangle\right)$.
We note that the clusters are not necessarily contiguous subsets of original spins of the chain i.e. there may be other clusters enclaved between constituent spins of a given cluster. 

The model in Eq. (\ref{H}) has a $Z_2$ symmetry as $\mathcal{H}(t)$ commutes with the parity operator $\hat{P}=\otimes_n\sigma_n^z$, which has eigenvalues $\pm 1$. As the annealing starts from the ground state of $\mathcal{H}(0)$, which is in the $+1$ sector, the time-dependent state must remain in the same sector throughout the process. The parity of the ground state is captured correctly by the SDRG approximation as the ground state $|\uparrow_{C_n}\rangle$ of each cluster has parity $+1$. As opposed to this, the first excited state of a cluster $|\downarrow_{C_n}\rangle=\frac{1}{\sqrt{2}}\left(\otimes_{i\in C_n}|\rightarrow_i\rangle - \otimes_{i\in C_n}|\leftarrow_i\rangle\right)$ has parity $-1$.  

Let us inspect how the SDRG cluster set of the instantaneous Hamiltonian depends on $t$, which is, for the time being, regarded as a control parameter. 
In the simple case of three spins, this is illustrated in the bottom row of Fig \ref{scheme}.
At $t=-\tau$, the initial Hamiltonian is at the edge of the disordered GM phase, and the clusters are simply the original spins. Thus, within the SDRG approximation, the ground state is a product state $\otimes_n|\uparrow_n\rangle$ in terms of the original spin variables. As opposed to this, at $t=\tau$, all fields are zero, so there will be only one ferromagnetic cluster comprising all spins of the chain. In between, for $-\tau<t<\tau$, the number density of clusters varies with $t$ from $1$ to $0$. It is not difficult to show the following monotonicity property of the cluster set. Let us assume that, for a given set of bonds $\{J_i\}$ and fields $\{h_i\}$, two given spins belong to a common cluster. Then, for a different set of bonds $\{J_i'\}$ and fields $\{h_i'\}$ fulfilling  $J_i'\ge J_i$ and $h_i'\le h_i$ for all $i$, the two spins must still belong to a common cluster.
For whichever annealing protocol, this has the following consequence. When monitoring the variation of the instantaneous SDRG cluster set as $t$ sweeps through the interval $[-\tau,\tau]$, we will observe a gradual aggregation of clusters starting from the completely disconnected set to the completely connected one, and splitting of existing clusters never occurs.

In the sequel, we will refer to the variation of the cluster set with the control parameter $t$ as the $t$-history of SDRG, while the aggregation-elimination process of spin clusters realized by the SDRG procedure when the energy scale  $\Omega$ is reduced from $1$ to $0$ is termed as the $\Omega$-history.   
With the $t$-history, we can associate a graph the nodes of which are the original spins and, at each fusion of a pair of spin clusters, a new link is added connecting the two corresponding constituent spins of fusing clusters. In this way, at the end of the $t$-history, a connected tree, called the annealing tree is produced. This contains short links connecting neighboring sites as well as long links connecting farther sites, the latter being as many as the missing links between adjacent sites (see Fig. \ref{tree}).

We conclude that the SDRG procedure provides us an effective Hamiltonian of the model at a fixed control parameter in the form
\be
\mathcal{H}_{\rm eff}(t)=-\sum_{i}\tilde h_{i}(t)\tilde\sigma^z_i,
\label{Heff}
\ee
where the summation goes over the set of spin clusters, and $\tilde\sigma_i$ is an effective spin associated with cluster $C_i$ acting on the space spanned by
$|\uparrow_{C_i}\rangle$ and $|\downarrow_{C_i}\rangle$.  
An improved approximation of the effective Hamiltonian may be obtained by extending Eq. (\ref{Heff})  with interaction terms of the form
\be
\mathcal{H}_{\rm int}(t)=-\sum_{ij}\tilde J_{ij}(t)\tilde\sigma^x_i\tilde\sigma^x_i.
\label{Hint}
\ee
Here, the effective couplings of a given cluster (with two other clusters) are obtained during the $\Omega$-history by recording the couplings of the cluster when it is eliminated by a field decimation.  
Eq. (\ref{Heff}) alone is a good approximation if the effective couplings are small compared to effective fields. This is typically valid except in the vicinity of a fusion event i.e., when, by driving the control parameter, a coupling becomes comparable with one of its neighboring fields. Beyond this point, the two involved clusters are unified to a common cluster, and we are back to an effective Hamiltonian given in  Eq. (\ref{Heff}) with one less quasi-independent effective spins.   

The above picture implies that the ground state of a cluster $C_n$ of spins is essentially unchanged until it merges with another cluster $C_m$. In that case, their state $|\uparrow_{C_n}\uparrow_{C_m}\rangle$ turns into  $|\uparrow_{C_n\cup C_m}\rangle\equiv\frac{1}{\sqrt{2}}(|\rightarrow_{C_n}\rightarrow_{C_m}+|\leftarrow_{C_n}\leftarrow_{C_m}\rangle)$. These fusion events are the points where the system is most exposed to defect formation during quantum annealing.
The reason for this is that, low-energy excitations of individual clusters from their ground state $|\uparrow_{C_n}\rangle$ to $|\downarrow_{C_n}\rangle$ are not allowed by conservation of parity. Although coherent excitations of any pairs of clusters, which can still have an arbitrarily low excitation energy, are not forbidden by parity symmetry, yet these have a low probability since the interaction between a typical pair of clusters is rather weak.
A considerable chance of excitation arises only when the effective coupling between two clusters is comparable with their effective transverse fields. This occurs at the point when the clusters are unified, i.e. when a new cluster is {\it in statu nascendi}.

\section{Elementary events of defect formation}
\label{fusion}

The variation of the ground state as follows from the effective Hamiltonian in Eq. (\ref{Heff}) is discontinuous: if the effective field $\tilde h_1$ of a cluster is infinitesimally above  or below one of its bonds $\tilde J$, it will be eliminated or joined to its neighboring cluster through bond $\tilde J$  at the scale $\Omega=\tilde h_1=\tilde J$, respectively, and the subsequent $\Omega$-history is the same in both cases.     
Now, focusing on just two clusters which are about to fuse, we refine this picture by including also the interaction term in Eq. (\ref{Hint}). 
We neglect the interaction with the rest of the system and retain only a two-dimensional space of cluster states spanned by the ferromagnetic state $|\uparrow_{n}\rangle$ (of parity $+1$) and $|\downarrow_{n}\rangle$ (of parity $-1$) of each cluster.
The effective Hamiltonian of this subsystem is
\be
\mathcal{H}_{\rm 12}(t)=-\tilde J(t)\tilde\sigma_1^x\tilde\sigma_2^x-\tilde h_1(t)\tilde\sigma_1^z -\tilde h_2(t)\tilde\sigma_2^z.
\label{H_eff}
\ee
As the interaction with the rest of the system is negligible, the state of the subsystem remains in the $+1$ parity sector spanned by the eigenvectors of $\mathcal{H}_{\rm 12}(t)$
\beqn
|E_-\rangle=\frac{1}{\sqrt{1+\alpha^2}}(|\uparrow_1\uparrow_2\rangle+\alpha |\downarrow_1\downarrow_2\rangle), \nonumber \\
|E_+\rangle=\frac{1}{\sqrt{1+\alpha^2}}(\alpha|\uparrow_1\uparrow_2\rangle-|\downarrow_1\downarrow_2\rangle).
\eeqn
where $\alpha(t)=\sqrt{1+r^2(t)}-r(t)$ and
\be 
r(t)=\frac{\tilde h_1(t)+\tilde h_2(t)}{\tilde J(t)}.
\ee
The corresponding energies are $E_{\mp}=\mp \sqrt{[\tilde J(t)]^2+[\tilde h_1(t)+\tilde h_2(t)]^2}=\mp\tilde J(t)\sqrt{1+r^2(t)}$. Since the effective bond (fields) monotonically increases (decrease) with $t$, there must be a point $t=t_0$ at which the energy gap $\epsilon(t)=E_+(t)-E_-(t)$ is minimal.  
As it follows from the structure of SDRG decimation rules, the renormalized couplings are products of a contiguous set of original couplings as
$\tilde J(t)=J_{i}(t)\prod_{j=1}^{l_b}\frac{J_{i+j}(t)}{h_{i+j}(t)}$ and  
$\tilde h_n(t)=h_{i_n+l_n}(t)\prod_{j=0}^{l_n-1}\frac{h_{i_n+j}(t)}{J_{i_n+j}(t)}$ for $n=1,2$. Here, $l_b$, $l_1$, and $l_2$ are the length of the effective bond and of the two clusters, respectively. 
Now we anticipate that defects forming at slow annealing rates are associated with fusion events of large clusters, which take place close to the critical point, so we may assume that $l_b,l_1,l_2\gg 1$.     
In this case, the dependence of couplings on $t$ can be written as
\beqn
\tilde J(t)\approx e^{-\mathcal{L}_b\delta(t)}\tilde J(t_0), \nonumber \\
\tilde h_n(t)\approx e^{\mathcal{L}_n\delta(t)}\tilde h_n(t_0),
\label{shift}
\eeqn
where the form of the generalized lengths $\mathcal{L}_b$ and $\mathcal{L}_i$, and the local reduced control parameter $\delta(t)$ of the fusion depend on the type of annealing protocol. For the linear one, the generalized lengths are just equal to the lengths $l_b$ and $l_i$, and
$\delta(t)=\ln\frac{p(t_0)}{p(t)}$, with $p(t)$ defined in Eq. (\ref{pt}).
For the special protocol, $\delta(t)=p(t_0)-p(t)$, while the generalized lengths are of the form $\sum_i\ln h_i^{-1}$, where the summation is over the original fields contained in the product expansion of $\tilde J(t)$ and $\tilde h_i(t)$, respectively. Thus, $\mathcal{L}_b$ and $\mathcal{L}_i$ are proportional to the real lengths $\mathcal{L}_b\sim l_b$ and $\mathcal{L}_i\sim l_i$ for large lengths, the proportionality constant tending to the average value of $\ln h_i^{-1}$.

At the point $t=t_0$, where the energy gap is minimal i.e. $\frac{d\epsilon}{d\delta}=0$, we have 
\be
\mathcal{L}_b\tilde J^2(t_0)\approx [\tilde h_1(t_0)+\tilde h_2(t_0)][\mathcal{L}_1\tilde h_1(t_0)+\mathcal{L}_2\tilde h_2(t_0)].
\label{emin}
\ee
As the fusion events relevant for large $\tau$ take place close to the critical point, where the distribution of fields is broadening on a logarithmic scale with decreasing $\Omega$ during the SDRG procedure \cite{fisher}, we may assume that $\tilde h_1(t_0)\gg \tilde h_2(t_0)$. 
This simplifies Eq. (\ref{emin}) to  $r(t_0)\approx \frac{\tilde h_1(t_0)}{\tilde J(t_0)}\approx \sqrt{\mathcal{L}_b/\mathcal{L}_1}$, where the ratio $\mathcal{L}_b/\mathcal{L}_1$ has an energy-scale independent limit distribution at large length scales \cite{fisher}. In terms of logarithmic couplings, we have $\ln\tilde h_1(t_0)-\ln\tilde J(t_0)\approx\frac{1}{2}\ln(\mathcal{L}_b/\mathcal{L}_1)$ which is an exponentially bounded $O(1)$ random variable.
Thus, we conclude that the point of fusion determined by the condition $\tilde J=\tilde h_1$ approximately coincides with the point $t_0$ at which the energy gap in the two-cluster problem is minimal, the difference becoming irrelevant at large length scales.  

Now, we can determine the crossover scale of annealing rate $\tau_c^{-1}$, which separates adiabatic fusions occurring for $\tau\gg \tau_c$ from completely diabatic ones ($\tau\ll \tau_c$). This can be achieved in two ways, both leading to the same conclusion.
First, we can invoke the adiabatic theorem, according to which the system will remain in the ground state of the instantaneous Hamiltonian as long as the annealing time satisfies
\be
\tau\gg\max_{0\le s\le 1}\frac{\left|\langle E_+(s)|\frac{d{\cal H}_{12}(s)}{ds}|E_-(s)\rangle\right|}{\epsilon^2(s)},
\label{at}
\ee
where $s\equiv\frac{t}{\tau}$ is a dimensionless time.
The matrix element in the numerator,  
$M(t)\approx \frac{\tilde J}{\sqrt{1+r^2}}[\mathcal{L}_br+\frac{\mathcal{L}_1\tilde h_1+\mathcal{L}_2\tilde h_2}{\tilde J}]|\frac{d\delta}{ds}|$, has a maximum at $r\sim O(1)$, which does not precisely coincide with the minimum position $t_0$ of the energy gap. Nevertheless, if we are interested in an estimate of the crossover annealing rate, we do not make a big mistake by evaluating it at $t=t_0$, where Eq. (\ref{emin}) holds.
This results in $M(t_0)\approx \frac{1}{2r(t_0)}\mathcal{L}_b\epsilon(t_0)|\frac{d\delta}{ds}|$, and we obtain for the crossover annealing rate,
$\tau_c^{-1}\sim \epsilon^2(t_0)/M(t_0)$, the following relation 
\be
\tau_c^{-1}\sim \epsilon(t_0)/\mathcal{L}_b.
\label{tauc}
\ee
The other way is to relate the problem in the crossover zone to the idealized two-level case solved by Landau and Zener, for which the probability of losing adiabaticity is
\be 
P_{\rm LZ}=e^{-2\pi\epsilon^2(t_0)/v},
\label{PLZ}
\ee
where $v$ is the slope of the linear asymptotes of approaching energy levels. Latter can be found by writing the Hamiltonian in the basis $\frac{1}{\sqrt{2}}(|E_-(t_0)\rangle\pm |E_+(t_0))$, in which the diagonal elements $e_{1}(t)$ and $e_{2}(t)=-e_{1}(t)$ vanish at $t=t_0$. Then $v=|\frac{d}{dt}(e_1-e_2)|_{t=t_0}|=\epsilon(r_0)\mathcal{L}_br_0^{-1}|\frac{d\delta}{ds}(t_0)|\tau^{-1}$.
This leads again to the scaling of the crossover annealing rate given in Eq. (\ref{tauc}).

Thus, beside that the crossover annealing rate is proportional to the minimal energy gap, it is also inversely proportional to the length of the connecting bond. The reason for this is that the renormalized couplings involved in the fusion process are products of original couplings, $O(l_b)$ in number. As a consequence, the effective annealing rate of a given fusion event is increased by a factor of $O(l_b)$ compared to the ``bare'' annealing rate $\tau^{-1}$. 

Now, let us consider what is the end state of the joint cluster well beyond the fusion for different values of $\tau$. 
For slow enough annealing rates, $\tau^{-1}\ll \epsilon(t_0)/\mathcal{L}_b$, the fusion of two clusters is flawless, the state following the instantaneous ground state, which can be also written as
$|E_-(t)\rangle=\frac{1}{\sqrt{2(1+\alpha^2)}}[(1+\alpha)|\rm FM\rangle+(1-\alpha)|\rm AF\rangle]$, where 
$|\rm FM\rangle=\frac{1}{\sqrt{2}}(|\rightarrow_1\rightarrow_2\rangle+|\leftarrow_1\leftarrow_2\rangle)$ and $|\rm AF\rangle=\frac{1}{\sqrt{2}}(|\rightarrow_1\leftarrow_2\rangle+|\leftarrow_1\rightarrow_2\rangle)$ are the ferromagnetic and antiferromagnetic (or domain wall) states of the block, respectively. The ground state, which is a combination of these states with equal amplitudes well before the fusion ($\alpha\approx 0$), $|E_-(0)\rangle=\frac{1}{\sqrt{2}}(|\rm FM\rangle+|\rm AF\rangle)$, will turn  into the state $|{\rm FM}\rangle$ well beyond the fusion when the coupling dominates over transverse fields ($\alpha\approx 1$).
Otherwise, for faster annealing, the fusion is imperfect and 
the state of the block will come into an oscillating combination of states $|\rm FM\rangle$ and $|\rm AF\rangle$ even well beyond the fusion ($r\approx 0$). In the extreme case, $\tau^{-1}\gg \epsilon(t_0)/\mathcal{L}_b$, the state essentially freezes to the initial state, thus the fusion can be regarded as a sudden quench, and we have at the end an oscillating combination $a_{\rm FM}(t)|{\rm FM}\rangle+a_{\rm AF}(t)|{\rm AF}\rangle$ with approximately equal amplitudes $|a_{\rm FM}|\approx |a_{\rm AF}|\approx \frac{1}{\sqrt{2}}$.

\section{SDRG picture of non-adiabatic annealing}
\label{nonadiabatic}

We argued in Sec. \ref{sdrg} that the annealing process, provided it is adiabatic, is realized by consecutive local fusions of strongly coupled spin clusters. Between fusion events, the system is described by a series of effective Hamiltonians of quasi-independent cluster spins. 
The question is how this picture is affected if some fusion events are imperfect, leaving the fusing clusters in a combination of ferromagnetically and antiferromagnetically coupled cluster spins. We will argue, that such defective fusions do not alter the cluster set i.e. the $t$-history will be the same as in the adiabatic case. The reason for this is that, even for an imperfect fusion, the newly born cluster will act as a whole when interacting with the rest of the system.  After an imperfect fusion occurred, the state of the new cluster is a combination of $|\rm FM\rangle$ and $|\rm AF\rangle$ states. The state of the total system at a later time is obtained as the combination of the state evolved from the initial state $|\rm FM\rangle$ of the new cluster and that evolved from $|\rm AF\rangle$.
Thus, at this point of the $t$-history, the further description of the system splits to two branches governed by different effective Hamiltonians, in the same manner as happens in the dynamical RSRG-t method \cite{va,monthus}. In the case when the new cluster is in state $|\rm FM\rangle$, the effective Hamiltonian will be the same as in the adiabatic case, given in Eq. (\ref{Heff}). However, if the new cluster is in state $|\rm AF\rangle$, then for constructing the effective Hamiltonian, a generalization of the standard SDRG procedure, the RSRG-X method \cite{pekker} has to be used. This method was developed to construct excited states of the RTIC, by using block reductions similar to that of the SDRG method with the difference that the local Hilbert space is restricted to the high-energy branch of states rather than to the low-energy one. For the RTIC, this modification affects exclusively the sign of effective couplings, otherwise the $\Omega$-history and the resulting cluster set is the same as for the standard SDRG. 
In the case of a bond decimation, the high-energy states
$|\uparrow_{12}\rangle\equiv\frac{1}{\sqrt{2}}(|\rightarrow_1\leftarrow_2\rangle+|\leftarrow_1\rightarrow_2\rangle)$ and $|\downarrow_{12}\rangle\equiv\frac{1}{\sqrt{2}}(|\rightarrow_1\leftarrow_2\rangle-|\leftarrow_1\rightarrow_2\rangle)$ are kept, and an effective term is generated $-\tilde h\tilde\sigma^z$ as in the case of the ground state branch, but with a different sign of the effective transverse field, $\tilde h=-\frac{h_1h_{2}}{J_1}$.

To illustrate the rules for fusions of clusters with anti-aligned states, 
let us consider an adiabatic fusion of a two-spin cluster having state 
$\frac{1}{\sqrt{2}}(|\rightarrow_1\leftarrow_2\rangle+|\leftarrow_1\rightarrow_2\rangle)$ with a one-spin cluster in state $\frac{1}{\sqrt{2}}(|\rightarrow_3\rangle+|\leftarrow_3\rangle)$. Then the post-fusion state depends on through which constituent spins the clusters are connected, as these spins will be aligned in the post-fusion state (in case of adiabatic fusion).    
Thus, if the clusters are connected through the bond between spin $1$ and $3$  then the post-fusion state  will be  $\frac{1}{\sqrt{2}}(|\rightarrow_1\leftarrow_2\rightarrow_3\rangle+|\leftarrow_1\rightarrow_2\leftarrow_3\rangle)$, whereas in case of a connection through spin $2$ and $3$ we have
$\frac{1}{\sqrt{2}}(|\rightarrow_1\leftarrow_2\leftarrow_3\rangle+|\leftarrow_1\rightarrow_2\rightarrow_3\rangle)$.
We can see that the rule for constructing the post-fusion state can be formulated without referring to the sign of effective fields of clusters. 
The only case when the sign of the field would affect a fusion event occurring later is when a fusion is realized through a bond which is over a defective cluster. For the sake of simplicity, we assume that it is a two-spin cluster. Now, in the branch in which the defective cluster is in state $|{\rm FM}\rangle$, the effective coupling over it is positive, and, consequently, the outer fusion will be realized by an aligned configuration of connected spins. 
In the branch, however, in which the internal cluster is in state $|{\rm AF}\rangle$, the effective bond over it is negative, forcing an outer fusion with anti-aligned states.
However, one can show that a fusion through a bond over a defective cluster cannot be adiabatic, as it must take place at lower energy and higher length scale than the fusion of the internal (defective) cluster. As a consequence, the negative sign of the field is irrelevant since any further fusion over a defective cluster must also be imperfect.

We come to the conclusion that the $t$-history i.e. the variation of the cluster set with $t$, is the same irrespective of whether the fusions are adiabatic or not. This circumstance does affect, however, the state of the system, as illustrated in the above examples, the generalization of which is straightforward.
After an adiabatic fusion, the new cluster comes to a state in which the original spins through which the two clusters are connected are aligned, whereas for a non-adiabatic fusion the post-fusion state is a combination of aligned and anti-aligned states.
Formally, the pre-fusion states of clusters $C_1$ and $C_2$ can be written as
$\frac{1}{\sqrt{2}}(1+\hat{P})\sum_ka_{k,C_1}|\dots\rightarrow_i\dots\rangle_{k,C_1}$ and $\frac{1}{\sqrt{2}}(1+\hat{P})\sum_la_{l,C_2}|\dots\rightarrow_j\dots\rangle_{l,C_2}$, respectively, where the states of spins $i\in C_1$ and $j\in C_2$ through which the two clusters are connected are displayed. 
Here, the effect of the parity operator $\hat{P}=\otimes_n\sigma_n^z$ is to flip the states polarized in the $x$-direction.  
In the case of a perfectly adiabatic fusion, the post-fusion state of the unified cluster will be 
$|{\rm FM}_{ij}\rangle=\frac{1}{\sqrt{2}}(1+\hat{P})\sum_{kl}a_{k,C_1}a_{l,C_2}|\dots\rightarrow_i\dots\rangle_{k,C_1}\otimes|\dots\rightarrow_j\dots\rangle_{l,C_2}$,
whereas, for a non-adiabatic fusion, it will be $a_{\rm FM}|{\rm FM}_{ij}\rangle+a_{\rm AF}|{\rm AF}_{ij}\rangle$ with
$|{\rm AF}_{ij}\rangle=\frac{1}{\sqrt{2}}(1+\hat{P})\sum_{kl}a_{k,C_1}a_{l,C_2}|\dots\rightarrow_i\dots\rangle_{k,C_1}\otimes\hat{P}|\dots\rightarrow_j\dots\rangle_{l,C_2}$.
In the completely diabatic case, we have $|a_{\rm FM}|^2=|a_{\rm AF}|^2=\frac{1}{2}$. 

\begin{figure}[ht]
\includegraphics[width=8cm]{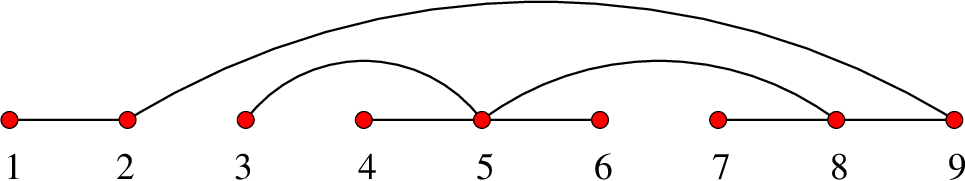}
\caption{\label{tree}
Illustration of the annealing tree. The long links corresponding to the missing short links $(2,3)$, $(3,4)$, and $(6,7)$ are, in order, $(2,9)$, $(3,5)$, and $(5,8)$. 
}
\end{figure}

In order to control quantum annealing one wishes to quantify how much the final state deviates from the true ground state. 
For the ferromagnetic Ising chain, a convenient measure of local errors is
\be 
N_{ij}=\frac{1-\langle\sigma_i^x\sigma_j^x\rangle}{2},
\ee
which is the probability of finding spins $i$ and $j$ in the (erroneous) anti-aligned state. In the case of an adiabatic fusion through the original spins $i$ and $j$, it takes the value $0$ in the post-fusion state, otherwise it is positive, and takes its maximal value $1/2$ if the fusion was completely diabatic.  
We remind the reader that the original spins $i$ and $j$ are not necessarily neighbors in the original lattice. Accordingly, the annealing tree associated with the $t$-history introduced in the Sec. \ref{sdrg} is not a linear graph but it has also long links connecting non-neighboring spins, and certain short links (between neighboring spins) are missing, see Fig. \ref{tree}.
There is, however, a one-to-one correspondence between missing short links and long links: for a given missing short link ($i,i+1$), the corresponding long link is the shortest link ($j,k$) which is ``over'' the missing short link i.e. the shortest link fulfilling $j\le i$ and $k\ge i+1$.
In the final state of the annealing process, we are interested in the defect density, $n=\overline{N_{i,i+1}}$, which is the average of local errors at short links. 
If a short link is part of the annealing tree, the contribution $N_{i,i+1}$ is directly available at the fusion of spin $i$ and $i+1$ in the $t$-history. For a missing short link, $N_{i,i+1}$ is determined by the post-fusion state of the cluster emerged by the fusion through the corresponding long link. It is easy to see that $N_{i,i+1}$ is predominantly determined by $N_{jk}$ of the corresponding long link. If $N_{jk}\approx0$ i.e. the fusion through the corresponding long link was adiabatic then the fusions through all other links below the long link must have been adiabatic, as well, consequently $N_{i,i+1}\approx 0$ i.e. the spins at the missing short link are aligned. If $N_{jk}>0$ and all other fusions along the shortest path from spin $i$ and $i+1$ are adiabatic then obviously $N_{i,i+1}=N_{jk}>0$. More generally, one can show that $N_{i,i+1}\ge N_{jk}$. This means that whenever a long link is defective ($N_{jk}>0$), the corresponding missing short link is also defective. In the extreme case of a completely diabatic fusion through a long link, we will have $N_{i,i+1}=1/2$ at the corresponding short link.

\section{The annealing gap}
\label{sec:gap}

We have seen that the excitations accessible for the annealing process are associated with fusion events of strongly coupled spin clusters. Now, we turn to the question what the joint probability density of the associated minimal gap and control parameter of these events is.

For the special annealing protocol introduced in Eqs. (\ref{Jdist}-\ref{pt}), this problem can be treated analytically.
In this case, the SDRG flow of coupling distributions is known not only at the critical point ($t=0$) by Fisher \cite{fisher}, but also in the entire disordered ($-\tau<t<0$) and ordered ($0<t<\tau$) GM phases by Igl\'oi \cite{igloi}. 
These are given by
\beqn
R(J,\Omega)=\frac{\theta_{\Gamma}}{\Omega}\left(\frac{J}{\Omega}\right)^{-1+\theta_{\Gamma}}, \nonumber \\
P(h,\Omega)=\frac{\pi_{\Gamma}}{\Omega}\left(\frac{h}{\Omega}\right)^{-1+\pi_{\Gamma}},
\label{solution}
\eeqn
where the running exponents $\theta_{\Gamma}\equiv R(\Omega,\Omega)\Omega$ and $\pi_{\Gamma}\equiv P(\Omega,\Omega)\Omega$ obey the flow equations
\be
\frac{d\theta_{\Gamma}}{d\Gamma}=\frac{d\pi_{\Gamma}}{d\Gamma}=-\theta_{\Gamma}\pi_{\Gamma}
\label{flow}
\ee
with the initial values $\theta_0$ and $\pi_0=\theta_0/p(t)$.
Here and in the sequel, we use the logarithmic energy scale $\Gamma\equiv-\ln\Omega$ rather than $\Omega$. 
Within the SDRG description, it is convenient to use the control parameter
\be
\Delta=\frac{\pi_0-\theta_0}{2}=-\frac{t}{\tau+t}\theta_0.
\ee
which is positive (negative) in the disordered (ordered) GM phase, and which is directly related to the dynamical exponent via $z=\frac{1}{2|\Delta|}$.
The solution of Eq. (\ref{flow}) for $\Delta\neq 0$ reads as \cite{igloi}
\be
\theta_{\Gamma}=\pi_{\Gamma}-2\Delta=\theta_0\Delta\frac{1-\tanh(\Gamma\Delta)}{\Delta+(\theta_0+\Delta)\tanh(\Gamma\Delta)},
\label{thetapi}
\ee
while, for $\Delta=0$, we have \cite{fisher}
\be
\theta_{\Gamma}=\pi_{\Gamma}=\frac{\theta_0}{1+\theta_0\Gamma}.
\ee
After having summarized the prerequisites for our analysis,
as a first step, we calculate the number density $Q(\Delta)$ of clusters, i.e. the ratio of the number of clusters produced in the $\Omega$-history to the number of original spins at a fixed $\Delta$.
We note that the growth of a cluster terminates (and the cluster is thereby reaches completion) when it is eliminated by a field decimation. Hence, $Q(\Delta)$ is simply the fraction of field decimations occurring in the entire $\Omega$-history from $\Gamma=0$ to $\Gamma=\infty$.
When the renormalization scale $\Gamma$ is infinitesimally increased to $\Gamma+d\Gamma$, a fraction $\pi_{\Gamma}d\Gamma$ of active (non-decimated) clusters is decimated. This gives a contribution $dQ=n_{\Gamma}\pi_{\Gamma}d\Gamma$ to $Q(\Delta)$, where $n_{\Gamma}$ is the ratio of the number of active clusters (or effective spins) at scale $\Gamma$ to the number of original spins. This is known to be
\be
n_{\Gamma}=\frac{1-\tanh^2(\Gamma\Delta)}{[1+(\theta_0/\Delta+1)\tanh(\Gamma\Delta)]^2}
\label{ngamma}
\ee
for $\Delta\neq 0$ \cite{igloi} and
\be
n_{\Gamma}=(1+\theta_0\Gamma)^{-2}
\label{ngamma0}
\ee
for $\Delta=0$ \cite{fisher}.
Using these results, we can evaluate the number density $Q(\Delta)=\int_0^{\infty}n_{\Gamma}\pi_{\Gamma}d\Gamma$, yielding
\be
Q(\Delta)=
\begin{cases}
1-\frac{1}{2}\frac{\theta_0}{\theta_0+2\Delta} &  (\Delta\ge 0) \\
\frac{1}{2}\frac{\theta_0+2\Delta}{\theta_0} &  (\Delta<0).
\end{cases}
\label{F}
\ee
We find that $Q$ decreases monotonically from $1$ to $0$ as the control parameter is varied from $\Delta=\infty$ ($t=-\tau$) to $\Delta=-\theta_0/2$ ($t=\tau$), taking the value $1/2$ at the self-dual critical point ($\Delta=t=0$). 
Knowing the function $Q(\Delta)$, we can easily obtain the density of fusion events occurring when the control parameter is infinitesimally changed from $\Delta$ to $\Delta+d\Delta$ as $\frac{dQ}{d\Delta}|d\Delta|$.

Next, we proceed to derive the conditional probability density of the energy scale of fusion events at a given control parameter.
Let us consider a fusion event which takes place when $\Delta$ is changed to $\Delta+d\Delta$, and denote the coupling between the fusing clusters by $\tilde J(\Delta)$, and the larger field by $\tilde h(\Delta)$. Then the $\Omega$-history at $\Delta$ and at $\Delta+d\Delta$ will differ in that, in the former case $\tilde h(\Delta)\gtrsim\tilde J(\Delta)$, and the cluster is eliminated by a field decimation at $\Omega=\tilde h(\Delta)$, while, in the latter case, $\tilde h(\Delta+d\Delta)\lesssim\tilde J(\Delta+d\Delta)$, and a fusion (bond decimation) occurs at $\Omega=\tilde J(\Delta+d\Delta)$. 
In the following, it is expedient to work with logarithmic couplings
$\zeta\equiv \ln\tilde J^{-1}$ and $\beta\equiv \ln\tilde h^{-1}$.
To obtain the infinitesimal shift of couplings with $\Delta$, we can use Eqs. (\ref{shift}), yielding 
$d\zeta \approx \mathcal{L}_b\frac{d\delta}{d\Delta}d\Delta$ and
$d\beta \approx -\mathcal{L}_1\frac{d\delta}{d\Delta}d\Delta$.
Now, at a fixed control parameter $\Delta$, let us consider a field decimation at some $\Gamma=\beta(\Delta)$, and one of the bonds of this cluster $\zeta(\Delta)$, for which obviously $\zeta(\Delta)>\beta(\Delta)$. The relative shift of these logarithmic couplings is $\frac{d\delta}{d\Delta}(\mathcal{L}_1+\mathcal{L}_b)|d\Delta|$.
The condition for a fusion to happen is that, at $\Delta+d\Delta$, the order of these couplings is reversed i.e. $\zeta(\Delta+d\Delta)<\beta(\Delta+d\Delta)$. 
This is equivalent to that $\zeta(\Delta)$ lies within the zone of width $\frac{d\delta}{d\Delta}(\mathcal{L}_1+\mathcal{L}_b)|d\Delta|$ near $\Gamma$.
As this condition contains couplings, as well as lengths of bonds and clusters, the further treatment of the problem would require the usage of joint distributions of couplings and lengths, which is difficult \cite{fisher,igloi}.
Instead we apply an approximation which greatly simplifies the problem. The distribution of lengths being narrow (exponential), we replace the variable $\mathcal{L}_1+\mathcal{L}_b\sim l_1+l_b$  by its average, which is inversely proportional to the fraction $n_{\Gamma}$ of active clusters at scale $\Gamma$.

The above considerations can be synthesized to obtain the conditional probability density of the energy scale $\Gamma$ of fusions at a fixed $\Delta$ as follows.
Changing the renormalization scale from $\Gamma$ to $\Gamma+d\Gamma$, an infinitesimal fraction $n_{\Gamma}\pi_{\Gamma}d\Gamma\equiv f_1(\Gamma,\Delta)d\Gamma$ of clusters (fields) is decimated. Then, such a field decimation will turn into a fusion at $\Delta+d\Delta$ if at least one of the bonds of the decimated cluster lies in the zone of width $\sim\frac{d\delta}{d\Delta}n_{\Gamma}^{-1}|d\Delta|$ near $\Gamma$. The probability for this is proportional to $2\frac{d\delta}{d\Delta}\theta_{\Gamma}n_{\Gamma}^{-1}|d\Delta|\equiv f_2(\Gamma,\Delta)|d\Delta|$.
Consequently, the conditional probability density of the energy scale $\Gamma$ of fusions at a fixed $\Delta$, which is proportional to $f_1f_2$, can be written in the form
\be
g(\Gamma|\Delta)d\Gamma=\mathcal{N}_{\Delta}\theta_{\Gamma}\pi_{\Gamma}d\Gamma,
\label{gdens}
\ee
where $n_{\Gamma}$ is luckily canceled and $\mathcal{N}_{\Delta}$ is a normalization constant.
From Eq. (\ref{flow}), we can see that $\theta_{\Gamma}\pi_{\Gamma}d\Gamma=-d\theta_{\Gamma}$, resulting in $g(\Gamma|\Delta)d\Gamma=-\mathcal{N}_{\Delta}d\theta_{\Gamma}$. The conditional distribution function $G_{>}(\Gamma|\Delta)=\int_{\Gamma}^{\infty}g(\Gamma'|\Delta)d\Gamma'$ is thus simply $G_{>}(\Gamma|\Delta)=\theta_{\Gamma}/\theta_0$ for $\Delta\ge 0$ and
$G_{>}(\Gamma|\Delta)=\pi_{\Gamma}/\pi_0$ for $\Delta<0$.
Using Eq. (\ref{thetapi}), we obtain 
\be 
G_{>}(\Gamma|\Delta)=
\frac{1-{\rm sgn}(\Delta)\tanh(\Gamma\Delta)}{1+(\theta_0/\Delta+1)\tanh(\Gamma\Delta)}
\label{dist}
\ee
for $\Delta\neq 0$, while we have
\be 
G_{>}(\Gamma|\Delta=0)=\frac{1}{1+\theta_0\Gamma}
\label{Gcrit}
\ee
at the critical point.
Finally, the joint probability density of the energy scale $\Gamma$ and control parameter $\Delta$ of fusions is of the form within the above approximation:
\be
g(\Gamma,\Delta)d\Gamma d\Delta=\mathcal{N}_{\Delta}\theta_{\Gamma}\pi_{\Gamma}\frac{dQ}{d\Delta}d\Gamma d\Delta,
\ee
where $\mathcal{N}_{\Delta}=1/\theta_0$ for $\Delta\ge 0$ and $\mathcal{N}_{\Delta}=1/\pi_0=1/(\theta_0+2\Delta)$ for $\Delta<0$. 

The distribution function in Eq. (\ref{dist}) outside of but close to the critical point ($\Delta\ll \theta_0$) changes its character at $\Gamma|\Delta|\sim O(1)$. For $\Gamma\ll |\Delta|^{-1}$, we have $G_{>}(\Gamma|\Delta)\approx G_{>}(\Gamma|\Delta=0)$, thus it is approximately equal to the critical distribution in Eq. (\ref{Gcrit}), whereas for $\Gamma\gg|\Delta|^{-1}$, it decays exponentially as
$G_{>}(\Gamma|\Delta)\simeq\frac{2|\Delta|}{\theta_0}e^{-2\Gamma|\Delta|}$.
This means that fusions are exponentially rare beyond a scale $\Gamma_{\Delta}=|\Delta|^{-1}$. Thus, the excitations which are effectively accessible during the annealing process have a finite energy gap in the GM phases, which closes according to
\be
\epsilon_{\Delta}\sim e^{-\frac{C}{|\Delta|}}
\label{gap}
\ee
as the critical point is approached ($|\Delta|\to 0$), $C$ denoting a constant. Note that, as opposed to the usual off-critical energy gap of homogeneous systems, the annealing gap is not a strictly forbidden gap. It merely expresses that the available energy levels are exponentially rare below the energy scale $\epsilon_{\Delta}$.
The energy $\epsilon_{\Delta}$ also marks out a crossover scale of the SDRG flow, above which ($\Omega\gg\epsilon_{\Delta})$ the renormalization trajectory is close-to-critical, while below this scale ($\Omega\ll\epsilon_{\Delta}$), the trajectory significantly departures from the critical one. In the former (critical) region, bond and field decimations occur on an equal rate, as $\theta_{\Gamma}\sim\pi_{\Gamma}\sim\Gamma^{-1}$, yielding that the probability of fusion events decays slowly with $\Gamma$, see Eq. (\ref{gdens}). If, however, $\Omega\ll\epsilon_{\Delta}$, either bond decimations (for $\Delta>0$) or field decimations (for $\Delta<0$) become rare [see Eq. (\ref{thetapi})], resulting in an exponential cutoff of fusion probability. 

The above considerations thus provide an a posteriori justification of the finiteness of gap outside of the critical point appeared implicitely in earlier works.  We find, however, that the vanishing of the annealing gap in Eq. (\ref{gap}) has a different functional form from the assumption in Eq. (\ref{epsilon_D}) \cite{dziarmaga_random}.

\section{Density of defects}
\label{defects}

\begin{figure}[ht]
\includegraphics[width=8cm]{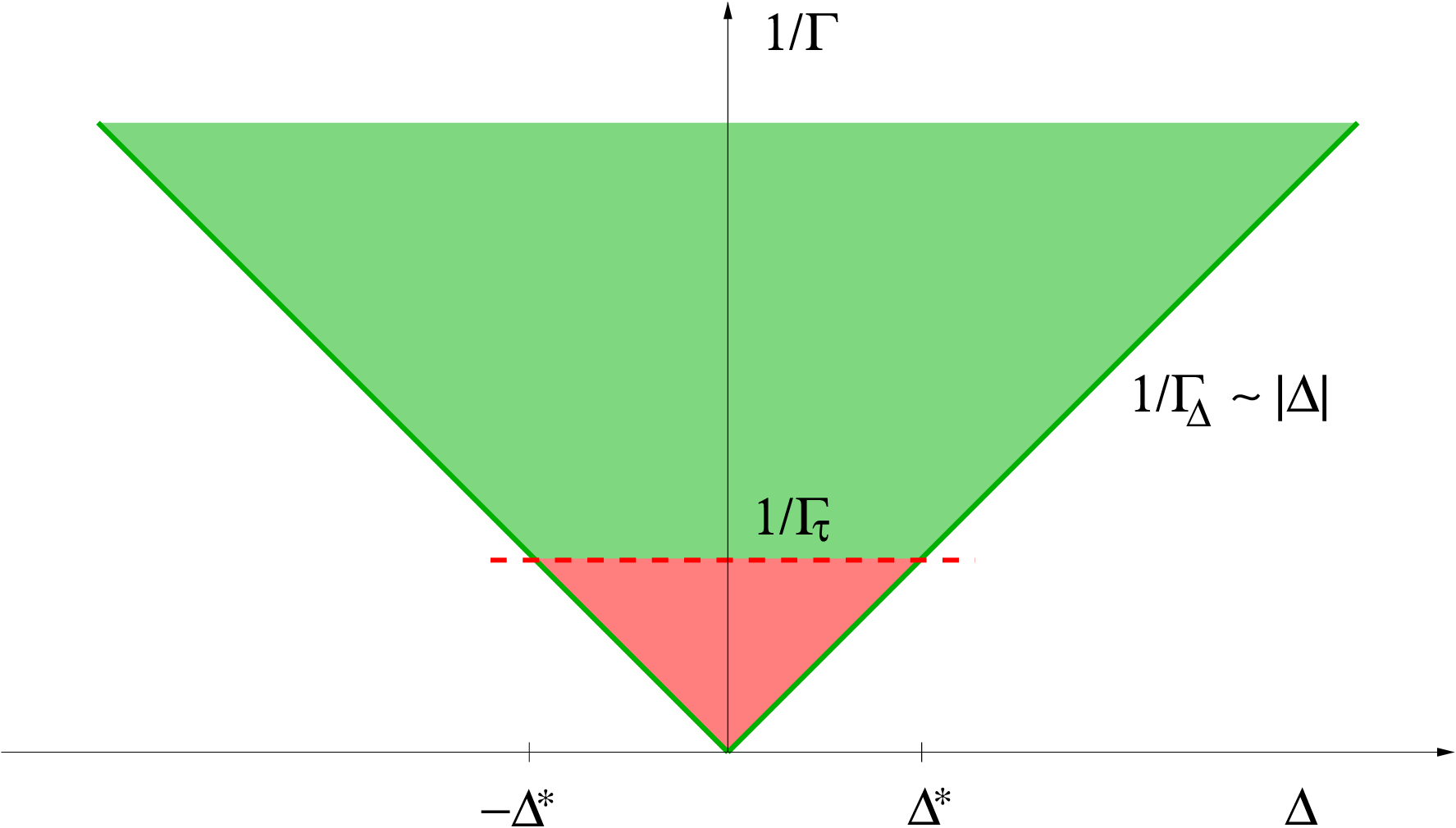}
\caption{\label{fig_gammadelta}
Schematic picture of fusion events in the $1/\Gamma$\textendash$\Delta$ plane. 
The green-shaded area shows the domain of fusion events which are flawless (adiabatic), while the erroneous (diabatic) fusions for a given annealing time $\tau$ are located in the pink-shaded area.
The boundary between the adiabatic and diabatic area indicated by the dashed red line is located at $1/\Gamma_{\tau}$ [see Eq. (\ref{gammatau}] and is shifted downwards with increasing $\tau$.
Note that the bounding lines of different areas symbolize exponential cutoff scales rather than strict boundaries. 
}
\end{figure}

After having established a model of annealing process in the RTIC, we proceed to an approximative calculation of the density of defects $n(t)$ produced from the beginning of the process up to time $t$.
This is proportional to the density of non-adiabatic fusions occurred from $t=-\tau$ ($\Delta=\infty$) up to time $t$, for which we may write
\be
n_{\tau}(t)\sim \int_{\Delta(t)}^{\infty}\int_0^{\infty}g(\Gamma,\Delta)P_{\rm LZ}d\Gamma d\Delta,
\label{nt}
\ee
where we used the alternative control parameter $\Delta(t)$ rather than $t$. 
Here, $P_{\rm LZ}$ is the Landau-Zener probability that a fusion is imperfect. As we obtained below Eq. (\ref{PLZ}), this can be written as $P_{\rm LZ}\sim e^{-C\epsilon(r_0)\mathcal{L}_b^{-1}\tau}$ with $C$ denoting a constant. Note that, as we found in the previous section, $g(\Gamma,\Delta)$ is non-negligible only within the critical zone $\Gamma\ll \Gamma_{\Delta}\sim|\Delta|^{-1}$, where the mean lengths of bonds and clusters are equal. Therefore we may stay within the approximation $\mathcal{L}_b\sim 1/n_{\Gamma}$ used in the derivation of $g(\Gamma,\Delta)$ in Sec. \ref{sec:gap}, and write
\be
P_{\rm LZ}(\Gamma)\sim e^{-C\Omega n_{\Gamma}\tau},
\label{LZ2}
\ee 
the dependence on $\Delta$ being negligible within the critical zone. 
We can see from Eq. (\ref{nt}) that the domain where there is a considerable amount of defect production is limited by two conditions. First, by $\Gamma\ll\Gamma_{\Delta}=|\Delta|^{-1}$ through the cutoff in the density function $g(\Gamma,\Delta)$. Second, within this domain, it is further limited by the cutoff in the Landau-Zener probability in Eq. (\ref{LZ2}) as $\Omega n_{\Gamma}\ll\tau^{-1}$. For large $\tau$, this condition can be explicitly written in terms of the energy scale by using Eq. (\ref{ngamma0}) as
\be
\Gamma\gg\Gamma_{\tau}\equiv\ln\frac{\tau/\tau_0}{\ln^2(\tau/\tau_0)},
\label{gammatau}
\ee
with some constant $\tau_0$. 
These two conditions mark out a domain of triangular shape in the $1/\Gamma$\textendash$\Delta$ plane, see Fig. \ref{fig_gammadelta}.
This means that essentially there is no defect formation
down to the value of the control parameter
\be
\Delta^*\sim \Gamma_{\tau}^{-1}\sim \left[\ln\frac{\tau/\tau_0}{\ln^2(\tau/\tau_0)}\right]^{-1},
\label{Dstar}
\ee
as well as beyond $-\Delta^*$ in the ferromagnetic phase. 

\begin{figure}[ht]
\includegraphics[width=8cm]{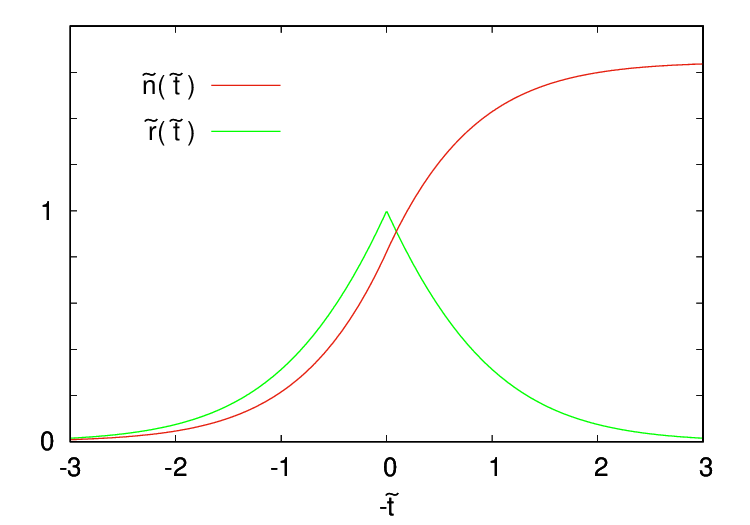}
\caption{\label{fig_scaling}
Scaling function of the time-dependent defect density $\tilde n(\tilde t)$ and that of the increase rate of defect density $\tilde r(\tilde t)$  for $\theta_0=1$. The scaling variable $\tilde t$ is defined in Eq. (\ref{ttilde}).  
}
\end{figure}

Now, we turn to the evaluation of the defect density in Eq. (\ref{nt}). A simple approximative form of $n(t)$ can be obtained by replacing the exponential cutoff dictated by $P_{\rm LZ}$ by a step function at $\Gamma_{\tau}$. 
This leads to
\be
n_{\tau}(t)\sim \int_{\Delta(t)}^{\infty}G_>(\Gamma_{\tau}|\Delta)\frac{dQ}{d\Delta}d\Delta.
\ee
Using Eqs. (\ref{F}) and (\ref{dist}), we find that the defect density has the scaling property
\be
n_{\tau}(t)=\frac{1}{\Gamma_{\tau}^2}\tilde n[\Gamma_{\tau}\Delta(t)],
\label{nscaling}
\ee
in the limit $\tau\to\infty$, in terms of the variable
\be 
\tilde t\equiv\Gamma_{\tau}\Delta(t)\simeq-\theta_0\frac{t}{\tau}\ln\frac{\tau}{\ln^2(\tau)}.
\label{ttilde}
\ee
The scaling function $\tilde n(\tilde t)$ is given as
\be
\tilde n(\tilde t)=\theta_0^{-2}\int_{\tilde t}^{\infty}\left(\frac{x}{\tanh(x)}-|x|\right)dx.
\label{scaling}
\ee
The increase rate of defect density $r_{\tau}(t)\equiv\frac{dn_{\tau}(t)}{dt}$ has a similar scaling property
\be
r_{\tau}(t)=\frac{1}{\Gamma_{\tau}\tau}\tilde r[\Gamma_{\tau}\Delta(t)],
\ee
with the scaling function $\tilde r(\tilde t)=\theta_0^{-1}(\tilde t/\tanh(\tilde t)-|\tilde t|)$. These scaling functions are plotted in Fig. \ref{fig_scaling}.
The scaling property of the time-dependent defect density in Eq. (\ref{nscaling}) is in agreement with the dynamical scaling hypothesis formulated in Ref. \cite{sadhukhan}. 
Eq. (\ref{nscaling}) implies that the defect density in the final state of the annealing process scales with $\tau$ as
\be 
n_{\tau}(\tau)\sim \left[\ln\frac{\tau/\tau_0}{\ln^2(\tau/\tau_0)}\right]^{-2}
\label{ntau}
\ee
for large $\tau$.

\section{Discussion}
\label{discussion}

In this work, we proposed a microscopic mechanism of defect formation in the quantum annealing of the random transverse-field Ising chain. This rests on the SDRG approach of the model and represents the change of the instantaneous Hamiltonian as an aggregation process of strongly coupled spin clusters. The clusters have a ferromagnetically ordered ground state, which is either preserved in pairwise fusions of clusters or broken depending on the effective annealing rate of fusion, the latter case leading to the appearance of defects in the final state. 
This picture leads to the conclusion that, although the Griffiths-McCoy phases surrounding the critical point are gapless, these phases are effectively gapped from the point of view of annealing. The reason for this lies in that the parity conservation of the model together with the weak interaction between clusters essentially restricts the possibility of defect formation to fusion events.
In the homogeneous variant of the model, the annealing process can be mapped to a set of Landau-Zener transitions in momentum space \cite{dziarmaga_prl}. In the disordered model, translational invariance is broken, but, according to our approach, the annealing process can still be decomposed (approximately) to a series of elementary two-level transitions, which are, however, irregular, being shaped by the fluctuations of disorder. 

This theory provides also quantitative predictions for the vanishing of the annealing energy gap, as well as for the scaling of defect density in the final state with $\tau$. Concerning the latter, our result given in Eq. (\ref{ntau}) contains a different form of logarithmic correction compared to Eq. (\ref{n_D}) found earlier by a simple adaptation of Kibble-Zurek scaling \cite{dziarmaga_random}. This difference originates in the correct identification and characterization of accessible excitations in the presence of disorder, which is beyond the capabilities of a simple scaling description.
To fit our result to Kibble-Zurek theory, we may regard $\Delta^*$ in Eq. (\ref{Dstar}), which is the edge of the domain within which defect production is considerable, as a freezing point. Indeed, the correlation length at this point, $\xi^*\sim (\Delta^*)^{-2}\sim \Gamma_{\tau}^{2}$, gives the inverse of defect density found in Eq. (\ref{ntau}). The usual condition of determining the freezing time, $1/\epsilon_{\Delta}\sim t^*$, with Eq. (\ref{gap}), provides correctly the leading term $(\ln\tau)^{-2}$ of the defect density but would give a different power of the inner logarithmic correction ($1$ instead of $2$).
To obtain the form found in Eq. (\ref{Dstar}), the freezing condition must be derived from the adiabatic theorem distilled into Eq. (\ref{tauc}), as
$\tau\sim\mathcal{L}_b/\epsilon_{\Delta}\sim \ln^2(1/\epsilon_{\Delta})/\epsilon_{\Delta}$. 

A further benefit of our theory is that it does not only predict the defect density at the end of annealing process but also provides an approximate form of the time-dependence of the density of defects accumulated up to time $t$. The corresponding scaling function is qualitatively similar to that found numerically by calculating the average occupation of instantaneous fermion modes as a function of time in Ref. \cite{sadhukhan}.  

Our quantitative results have been obtained by assuming a particular distribution of couplings and for a particular annealing protocol. Nevertheless, the $\Delta$-domain in which defect production is considerable shrinks to the critical point $\Delta=0$ in the limit $\tau\to\infty$, furthermore, the solution for the SDRG flow of distributions in Eq. (\ref{solution}), which we used in the analysis is known to be attractive for any (sufficiently regular) initial distribution of couplings \cite{fisher}. Therefore we expect our asymptotic results, such as the vanishing of the annealing gap in Eq. (\ref{gap}) or the scaling of defect density with $\tau$ in Eq. (\ref{ntau}) to be valid universally i.e. irrespective of the form of coupling distribution and the details of annealing protocol.

  The verification of the asymptotic scaling of defect density in Eq. (\ref{ntau}) is a hard task both by a numerical solution of the dynamics or experimentally, since, due to the slow logarithmic scaling, access to large time scales would be required. Recent numerically exact calculations \cite{caneva,zanca} show clear signs of the leading logarithmic behavior but the available time scales are insufficient to validate the form of the inner logarithmic correction. Promising alternatives are the experiments carried out with programmable superconducting annealers consisting of a few thousand qubits \cite{king,king2023}. At the present time, such tests are limited again by the available time scale since, owing to the onset of thermal effects, deviations from coherent behavior appear for longer anneals.     
 
Although the mechanism of defect formation presented in this work is formulated for the one-dimensional transverse-field Ising model, much of the theory can be generalized to higher dimensions.
The SDRG method can be applied also to higher dimensional variants of the model, with the difference that this technique is restricted there to a numerical implementation instead of an analytical treatment \cite{ki}. Nevertheless, this method predicts ground state properties qualitatively similar to those of the RTIC.
Therefore, the decomposition of the annealing process to a series of fusions of strongly coupled clusters is expected to hold generally for higher dimensional variants of the transverse-field Ising model, as well. The study of quantum annealing in such models is left for future research.

\begin{acknowledgments}
The author thanks F. Igl\'oi, I. A. Kov\'acs, and G. Ro\'osz for useful discussions.
This work was supported by the National Research, Development and Innovation Office NKFIH under Grant No. K146736 and by the Quantum Information National Laboratory of Hungary.
\end{acknowledgments}


\end{document}